\documentclass[twocolumn,showpacs,preprintnumbers,draftclsnofoot]{revtex4}
\usepackage{amsmath}
\usepackage{amssymb}
\usepackage{graphicx}
\usepackage{dcolumn}
\usepackage{bm}
\usepackage{amssymb}
\usepackage{graphicx}
\usepackage{dcolumn}
\usepackage{bm}
\begin{document}

\title{ Floquet Engineering of Two Dimensional Photonic Waveguide Arrays with $\pi$ or $\pm2\pi/3$ Corner states  }
\author{Ma Luo\footnote{Corresponding author:luoma@gpnu.edu.cn} }
\affiliation{School of Optoelectronic Engineering, Guangdong Polytechnic Normal University, Guangzhou 510665, China}

\begin{abstract}

In this paper, we theoretically study the Floquet engineering of two dimensional photonic waveguide arrays in three types of lattices: honeycomb lattice with Kekule distortion, breathing square lattice and breathing Kagome lattice. The Kekule distortion factor or the breathing factor in the corresponding lattice is periodically changed along the axial direction of the photonic waveguide with frequency $\omega$. Within certain ranges of $\omega$, the Floquet corner states in the Floquet band gap of quasi-energy spectrum are found, which are localized at the corner of the finite two-dimensional arrays. Due to particle-hole symmetric in the model of honeycomb and square lattice, the quasi-energy level of the Floquet $\pi$ corner states is $\pm\omega/2$. On the other hand, Kagome lattice does not have particle-hole symmetric, so that the quasi-energy level of the Floquet $\pm2\pi/3$ corner states is near to $\pm1\omega/3$. The corner states are either protected by crystalline symmetry or reflection symmetry. The finding of Floquet fractional-$\pi$ corner states could provide more options for engineering of on-chip photonic devices.

\end{abstract}

\maketitle

\section{Introduction}

Topological properties of physical lattice models are featured by robust edge states that exist in lower spatial dimension. Specifically, the second order topological insulator of two dimensional lattice model have robust corner modes at the corners of a finite flake \cite{Khalaf18,Ezawa181,Ezawa182,Guido18,Akishi18,Franca18}. Although the lattice models have been demonstrated to host topological corner states, the experimental observation of the corner state in condensed matter physical systems is challenging \cite{Weixuan21,Yuanfeng19}, because the lattice models require non-trivial hopping terms. The attempts to observe the topological corner states have been draw to other physical systems that mimic the topological lattice models, such as topological circuits \cite{Huanhuan20}, topological sonic crystals \cite{Huahui21} and photonic crystals \cite{BiYeXie18,BiYeXie19,XiaoDong19}.

In order to obtained the topological corner states on demand, Floquet engineering of the lattice models have been proposed \cite{TanayNag21,RuiXing21,Chaudhary20,Ghosh20b,Biao20,Haiping20,Tanay19,Ranjani19,Martin19,Raditya19}. Periodical perturbation of the hopping terms and (or) the on-site potential of the lattice model effectively change the Hamiltonian, which could drive the systems into topological phase. The topological feature of the corner modes have been enriched by introducing non-Hermitian terms into the Floquet systems \cite{HongWu21,Jiaxin20}. One of the most interesting application of Floquet second order topological phase is to generated corner states in topological superconductor \cite{Ghosh20,Ghosh20a,Bomantara20,Kirill19}, so that the Floquet Majorana corner modes could be applied for quantum computing physics \cite{Raditya20}.

Because of the similarity between the $Schr\ddot{o}dinger$ equations of quantum systems and the mode coupling equations of photonic waveguide arrays, the topological phase of two dimensional lattice models can be mimicked by finite photonic waveguide arrays \cite{Longhi03,Rechtsman13,Tomoki19,Ablowitz19,Leykam16,WeiweiZhu21} or photonic microring resonator lattices \cite{Shirin18}. Floquet engineering of one dimensional photonic waveguide arrays have been thoroughly studied in theory and experiment. Each waveguide is periodically curved, so that the coupling strength between the nearest neighboring waveguides are periodically modified. By arraying the waveguides with alternating interval, the Floquet Su-Schrieffer-Heeger (SSH) model can be mimicked. Theoretical study found that zero modes and $\pi$ modes, which are localized at the end of the one dimensional lattice, can be generated by the Floquet engineering \cite{BoZhu18,Kuzmiak20,Yiming20,Bisianov20,Shengjie21,LuQi21}. The localized modes have been observed in experiment \cite{Qingqing15,QingqingC19}. Extending to two dimensional photonic waveguide arrays, the first \cite{Ablowitz19,Leykam16,Shirin18} and second \cite{WeiweiZhu21} order Floquet topological insulating phases have been theoretically studied, but the experimental observation is limited to the first order Floquet topological insulating phase\cite{Sebabrata17,Lukas17}. The second order topological insulating phase with zero energy corner states is experimentally observed in two dimensional array of straight waveguide in Kagome lattice \cite{Kirsch21}. However, the Floquet $\pi$ mode with nonzero energy cannot be found in the systems without Floquet modulation.

By further engineering multiple types of two dimensional array of waveguide with Floquet modulation, novel type of topological corner states could be found. Thus, we theoretically studied photonic waveguide array in three types of lattice structure, which host localized corner modes. The proposed structure could be implemented in experiment by femtosecond laser direct writing techniques\cite{Davis96}. The photonic waveguides are arrayed in honeycomb lattice with Kekule distortion, breathing square lattice (which is similar to coupling multiple parallel 1D SSH chains with alternating strength), or breathing Kagome lattice. For the lattice arrays of straight waveguide, the static models of the three types of lattice have different properties. For the honeycomb lattice with static Kekule distortion, the two Dirac cones mix with each other, which opens a band gap. For the breathing square lattice, if the SSH lattice in each chain is topological, an isolated edge band is generated by coupling the end states of the 1D SSH chains. For the breathing Kagome lattice, the model could be in the second order topological insulating phase with zero energy corner states, depending on the breathing factor. As the waveguides being periodically curved, the coupling strength between neighboring waveguide is periodically changed along the axial direction. Effectively, the bulk bands, edge bands and corner states are modified. Within the Floquet band gap of the quasi-energy spectrum, anomalous types of $\pi$ and $\pm2\pi/3$ corner states could appear. The fractional-$\pi$ corner states is due to the presence of Floquet topological band gap in the absence of particle-hole symmetric. As comparison, the previously predicted fractional-$\pi$ corner states in Floquet systems are due to interaction \cite{Bomantara21,Bomantara21a}.

The paper is organized as follows: In Sec. II, the lattice structure and modulation scheme of the waveguide are described; the theoretical method is described. In Sec. III, the numerical results of the waveguide arrays in three types of lattice are presented and discussed. In Sec. IV, the discussion and conclusion is given.

\section{Theoretical model}

\subsection{Structure of the finite waveguide array in three types of lattices}

\begin{figure*}[tbp]
\scalebox{0.52}{\includegraphics{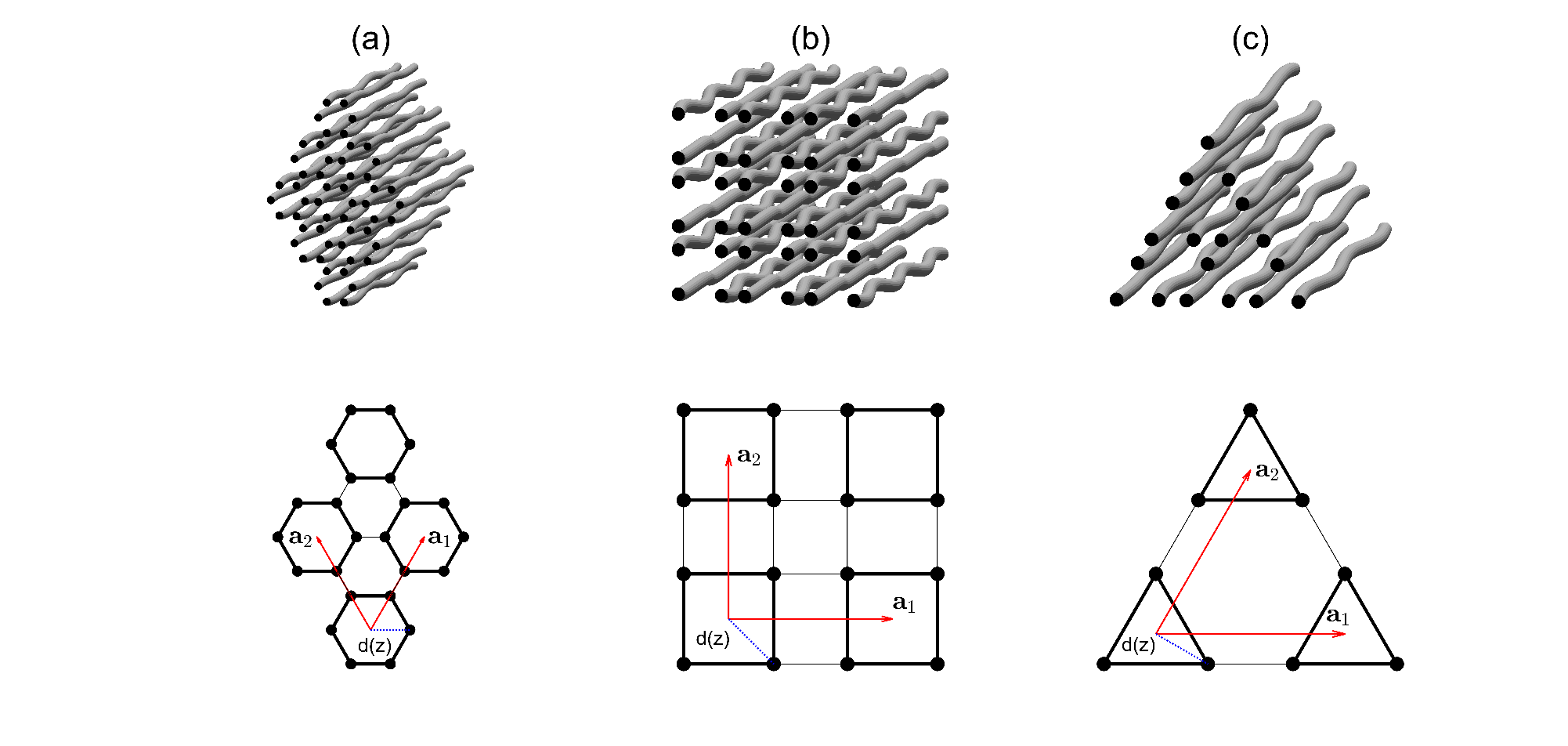}}
\caption{ The three dimensional view of the two dimensional arrays of periodically modulated circular optical waveguides in (a) hexagonal lattice with Kekule distortion, (b) breathing square lattice, (c) breathing Kagome lattice. The demonstration of the lattice structure of the hexagonal lattice with Kekule distortion, the breathing square lattice, and the breathing Kagome lattice are plotted in (d), (e), and (f), respectively. The nearest neighbor coupling between two waveguides within the same unit cell and in different unit cells are indicated as thick and thin solid (black) lines, respectively. The two basis vectors of the primitive unit cell of each lattice structure, designated as $\mathbf{a}_{1(2)}$, are plotted as red vectors. A finite array have $N$ unit cells along each basis vector of the primitive unit cell. The distance between each waveguide and the center of the corresponding unit cell, $d(z)$, is indicated by the dashed (blue) lines, which is periodic about the axial coordinate $z$.  }
\label{figure_system_kekule}
\end{figure*}

The three dimensional structure of the periodically modulated finite waveguide array in honeycomb lattice with O-type Kekule distortion, breathing square lattice, and breathing kagome lattice are plotted in Fig. \ref{figure_system_kekule}(a), (b), and (c), with the demonstration of the corresponding lattice structure in (d), (e), and (f), respectively. For each of the the three lattice structures, there is six, four, or three waveguide in each primitive unit cell, respectively. The distance between each waveguide and the center of the primitive unit cell is periodically modulated as a function of the axial coordinate, i.e. $d(z)=d_{0}+d_{1}+d_{2}\sin(2\pi z/L)$, where $d_{1}$ ($d_{2}$) is the static (dynamic) distortion or breathing factor and $L$ is the period of the modulation. The lattice constant of the three models are $3d_{0}$, $2\sqrt{2}d_{0}$, and $2\sqrt{3}d_{0}$, respectively. The finite array of the honeycomb lattice and the square lattice include $N\times N$ primitive unit cells. For the square lattice, $N$ could be either integer or an integer plus $\frac{1}{2}$. In case that $N$ is an integer plus $\frac{1}{2}$, two (three) of the four edges (corners) are consisted of half (quarter) unit cell. The finite array of the Kagome lattice include $N(N-1)/2$ primitive unit cells.

\subsection{Tight binding model}

The mode coupling theory is applied to describe the waveguide arrays. Only the coupling between the nearest neighboring waveguide is considered in the model. The coupling strength between two waveguide is an exponentially decay function of the separation between the two waveguide, i.e. $c(d_{i,j})=c_{0}^{\prime}e^{-d_{i,j}/\delta}$, where $c$ is the coupling strength with separation between the i-th and j-th waveguide being $d_{i,j}$, $c_{0}^{\prime}$ and $\delta$ are the structural parameters of the waveguide. For the nearest neighboring pair of waveguide in the same or different primitive unit cell, $d_{i,j}$ is proportional to $d(z)$ with a coefficient $r_{1}$ or $r_{2}$, respectively, which are dependent on the type of lattice. Thus, the mode coupling equations can be written in the form of $Schr\ddot{o}dinger$ equations, with the time being replaced by the axial coordinate $z$, and the hopping between the nearest neighboring sites has strength $c(d_{i,j})$ \cite{Shengjie21}. The Hamiltonian is given as $H=\sum_{\langle i,j\rangle}c(d_{i,j})a_{i}^{\dag}a_{j}$ with the summation covering the nearest neighboring sites, $a_{i}^{\dag}$ ($a_{i}$) being the creation (annihilation) operator of optical mode at the i-th waveguide. Since the coupling strength is periodic function of $z$ with period $L$, the Hamiltonian can be expanded as Fourier series of axially oscillating Hamiltonian with frequency being $n\omega$ where $\omega=2\pi/L$, and $n$ is a series of integer. Specifically, the coupling strength between two waveguide in the same (different) primitive unit cell is expanded as
\begin{equation}
c(d_{i,j})=c_{0}^{\prime}e^{-\frac{[d_{0}+d_{1}+d_{2}\sin(2\pi z/L)]r_{1(2)}}{\delta}}=\sum_{n}c_{n}e^{\frac{2\pi nz}{L}}
\end{equation}
where
\begin{equation}
c_{n}=\int_{0}^{L}{c_{0}^{\prime}e^{-\frac{[d_{0}+d_{1}+d_{2}\sin(2\pi z/L)]r_{1(2)}}{\delta}-\frac{i2\pi nz}{L}}dz}
\end{equation}
Thus, the Hamiltonian is expanded as $H(z)=\sum_{n=-\infty}^{+\infty}H_{n}e^{\frac{2\pi nz}{L}}$.

According to the Floquet theorem, the periodically varying Hamiltonian can be transformed into effective Floquet Hamiltonian $H_{F}$, which is axially-independent (independent of $z$) \cite{Shengjie21,DalLago15,Fedorova19,Thuberg16,Reyes17}. The effective Hamiltonian is consisted of diagonal block, which are $H_{n}$ with diagonal element being shifted by $n\omega$ for the $n$-th replica; non-diagonal block at the $n$-th row and $m$-th column as $H_{n-m}$. Diagonalization of the effective Hamiltonian gives the quasi-energy level $\varepsilon$ and the corresponding eigen-state $|\psi\rangle=e^{-i\varepsilon z}\sum_{n=-\infty}^{+\infty}\varphi_{i,n}e^{in\omega z}$, where $\varphi_{i,n}$ is the amplitude at the $i$-th waveguide of the $n$-th replica. For high frequency axially oscillation, the Floquet replica can be truncated as $|n|\leq n_{Max}$. The mode amplitude at the i-th waveguide (lattice site) is given as $\rho_{i}=\sum_{n=-n_{Max}}^{+n_{Max}}|\varphi_{i,n}|^{2}$. In our numerical simulation, the system with $\omega/c_{0}>2$ is considered, with $c_{0}=c_{0}^{\prime}e^{-d_{0}r_{1(2)}/\delta}$. For the static model with $d_{2}=0$, the band width of the investigated models is about $6c_{0}$ (i.e. the energy band ranges within $[-3c_{0},3c_{0}]$). For the corresponding dynamic system with $\omega/c_{0}\equiv\xi$, the energy band of the n-th replica is approximately within $[n\xi c_{0}-3c_{0},n\xi c_{0}+3c_{0}]$. In order to have accurate result within the quasi-energy range $\varepsilon\in[-\omega,\omega]\equiv[-\xi c_{0},\xi c_{0}]$, the replica with energy band overlapping with the energy range $[-\xi c_{0},\xi c_{0}]$ must be included in the effective Floquet Hamiltonian, i.e. for $n$ that satisfies $|n|\xi c_{0}-3c_{0}<\xi c_{0}$, the corresponding replica need to be kept. Since we focus on the systems with $\xi>2$, $n_{Max}=2$ is enough to have high accuracy. The inference is confirmed by the numerical results with $n_{Max}=3$ (not shown in this article), which have negligible different from those with $n_{Max}=2$.

\section{Numerical result}

The numerical result of waveguide arrays in three types of lattice structure is summarized in the following three subsections.

\subsection{Honeycomb lattice with Kekule distortion}

For the honeycomb lattice model with dynamical Kekule distortion, the tight binding Hamiltonian is given as
\begin{equation}
H_{n}=\begin{bmatrix}
0 & c_{n,1} & 0 & c_{n,2}^{(1)} & 0 & c_{n,1} \\
c_{n,1} & 0 & c_{n,1} & 0 & c_{n,2}^{(2)} & 0 \\
0 & c_{n,1} & 0 & c_{n,1} & 0 & c_{n,2}^{(3)} \\
c_{n,2}^{(1)*} & 0 & c_{n,1} & 0 & c_{n,1} & 0 \\
0 & c_{n,2}^{(2)*} & 0 & c_{n,1} & 0 & c_{n,1} \\
c_{n,1} & 0 & c_{n,2}^{(3)*} & 0 & c_{n,1} & 0
\end{bmatrix}
\end{equation}
where $c_{n,2}^{(1)}=c_{n,2}e^{i\mathbf{k}\cdot\mathbf{a}_{1}}$, $c_{n,2}^{(2)}=c_{n,2}e^{i\mathbf{k}\cdot\mathbf{a}_{2}}$, $c_{n,2}^{(3)}=c_{n,2}e^{i\mathbf{k}\cdot(\mathbf{a}_{2}-\mathbf{a}_{1})}$, $\mathbf{k}$ is the Bloch wave vector, $c_{n,1}=\int_{0}^{L}{c_{0}exp\{-\frac{[d_{1}+d_{2}\sin(2\pi z/L)]}{\delta}-\frac{i2\pi nz}{L}\}dz}$, and $c_{n,2}=\int_{0}^{L}{c_{0}exp\{-\frac{[-2d_{1}-2d_{2}\sin(2\pi z/L)]}{\delta}-\frac{i2\pi nz}{L}\}dz}$. In the numerical calculation, $d_{0}/\delta=3$ is assumed. The indices of lattice site in one unit cell is sorted by counter-clockwise order. Assuming $n_{Max}=2$, the effective Floquet Hamiltonian is given as
\begin{equation}
H_{eff}=\begin{bmatrix}
H_{0}-2\omega\mathbf{I} & H_{-1} & H_{-2} & H_{-3} & H_{-4} \\
H_{1} & H_{0}-\omega\mathbf{I} & H_{-1} & H_{-2} & H_{-3} \\
H_{2} & H_{1} & H_{0} & H_{-1} & H_{-2}  \\
H_{3} & H_{2} & H_{1} & H_{0}+\omega\mathbf{I} & H_{-1} \\
H_{4} & H_{3} & H_{2} & H_{1} & H_{0}+2\omega\mathbf{I}
\end{bmatrix} \label{effHamiHoney}
\end{equation}
where $\mathbf{I}$ is identity matrix with the same size as $H_{n}$. Because the bending of the waveguide is strictly along the direction that starts from the center of the unit cell and points to the corresponding lattice site, the modulation of the structures preserves the $C_{6}$ rotational symmetric of the honeycomb lattice with static Kekule distortion.

For the static model with $d_{2}=0$ and $d_{1}\ne0$, the bulk energy band is gapped. The zigzag nanoribbons have two nearly flat bands near to zero energy, but the armchair nanoribbons remain being gapped. We considered the nanoflake with armchair edge and four corners, including two zigzag corners and two armchair corners. When the static Kekule factor $d_{1}$ is positive, the coupling strength of intra-unit cell is weaker than that of inter-unit cell. A zero energy corner state appear at each zigzag corner. The honeycomb lattice with Kekula distortion has $C_{6}$ rotational symmetry. The presence of zero energy corner state is protected by the crystalline symmetric, so that the system is in the topological crystalline insulators (TCIs) phase \cite{Wladimir19}. The topological invariant of the band gap around zero energy is defined in  Ref \cite{Wladimir19} as $\chi^{(6)}=([M_{1}^{(2)}],[K_{1}^{(3)}])$, which equates to $(2,0)$. The corner charge, defined as $mod(\frac{1}{4}[M_{1}^{(2)}]+\frac{1}{6}[K_{1}^{(3)}],1)$, equates to $\frac{1}{2}$, so that there are two corner states. On the other hand, if $d_{1}$ is negative, the topological invariant is $\chi^{(6)}=(0,0)$, so that there is no zero energy corner state.

\begin{figure}[tbp]
\scalebox{0.68}{\includegraphics{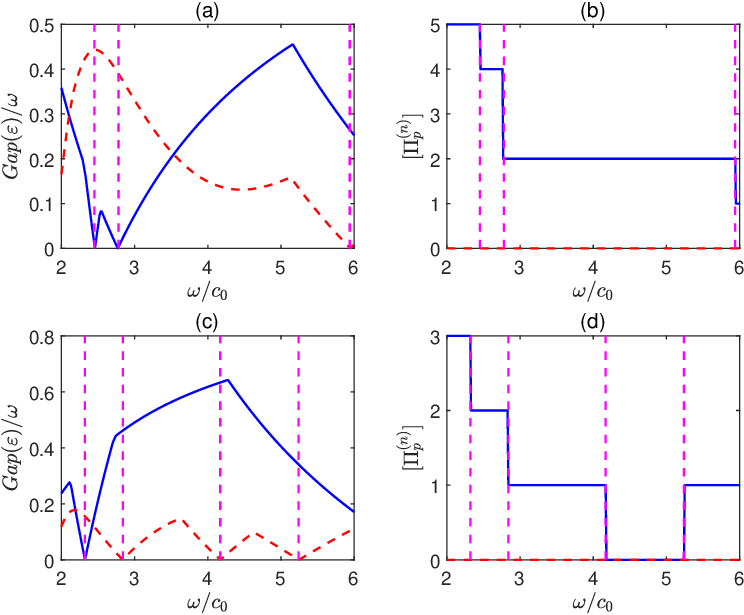}}
\caption{ For the model in Fig. \ref{figure_system_kekule}(a) with $d_{1}=0.1d_{0}$ and $d_{1}=-0.1d_{0}$, as functions of $\omega/c_{0}$, the bulk band gaps around quasi-energy level $\omega/2$ at high symmetric point of the first Brillouin zone are plotted in (a) and (c), respectively; the topological invariant of the band gap are plotted in (b) and (d), respectively. In (a,c), the band gap at the $\Gamma$ and $M$ points are plotted as blue(solid) and red(dashed) lines, respectively. In (b,d), the topological invariant $[M_{1}^{(2)}]$ and $[K_{1}^{(3)}]$ of the band gap are plotted as blue(solid) and red(dashed) lines, respectively. }
\label{figure_kekule_bulkTopo}
\end{figure}

\begin{figure*}[tbp]
\scalebox{0.74}{\includegraphics{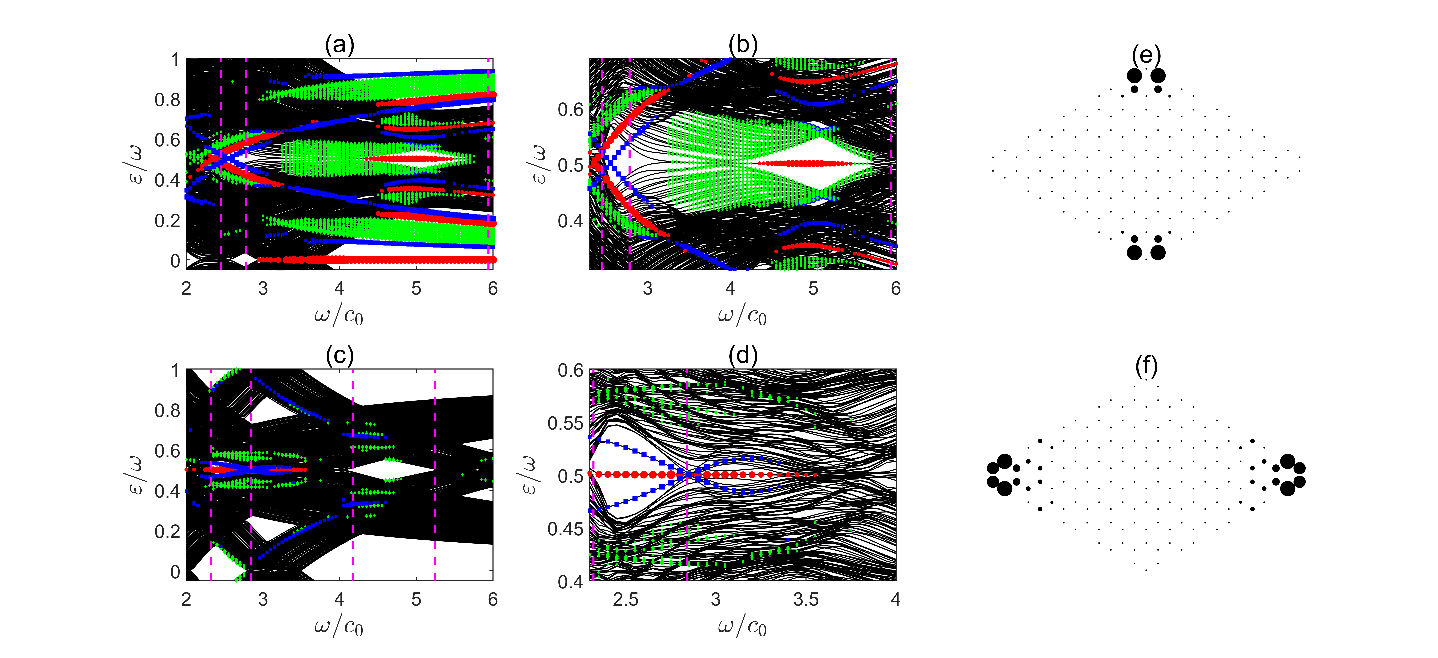}}
\caption{ (a-d) Quasi-energy spectrum of the Kekule distorted hexagonal lattice in Fig. \ref{figure_system_kekule}(a) with $N=11$ as a function of $\omega/c_{0}$. The static Kekule distortion factors for the systems in (a) and (c) are $d_{1}=0.1d_{0}$ and $d_{1}=-0.1d_{0}$, respectively. The oscillating Kekule distortion factors for both systems are $d_{2}=0.15d_{0}$. (b) and (d) are the zoom in view of (a) and (c) near to the spectrum of the $\pi$ modes, respectively. The states that are localized at the zigzag corners, armchair corners, and edges are marked by red (circle) dots, blue (square) dots, and greed (diamond) dots, respectively. The size of the markers is proportional to the degree of localization, i.e., $\sum_{i\in\Omega}\rho_{i}$. The vertical magenta dashed lines marked the $\omega$ where the bulk band gap at high symmetric points closes and the value of the topological invariant $\chi^{(6)}$ changes. (e) and (f) are the spatial distribution of the mode amplitude of the $\pi$ corner states and the armchair corner states, respectively. }
\label{figure_kekule_band}
\end{figure*}

In the presence of the dynamic modulation, the band structure as well as the topological properties are modified. In the low frequency limit, i.e., $\omega\rightarrow0$, the systems are in adiabatic pumping cycles, which can be described by the topological pump theory \cite{Wladimir22}. Because we consider the system with sizable value of $\omega$, the Floquet theory is applied to describe the systems. The presence of $\omega/2$ quasi-energy corner state can be explained by extending the definition of the topological invariant of the $C_{6}$ rotational symmetric, i.e., $\chi^{(6)}$, into the Floquet system described by the effective Floquet Hamiltonian in Eq. (\ref{effHamiHoney}). For the system with $C_{n}$ symmetric, the n-fold rotational operators act on each Floquet replica with the same symmetric matric. The eigenvalue of the n-fold rotational operators of the states at high symmetric points of the first Brillouin zone equate to $\Pi_{p}^{(n)}=e^{2\pi i(p-1)/n}$ with $p$ being integer in $[0,n-1]$, $\Pi$ representing the high symmetric point $\Gamma$, $M$, or $K$. For the honeycomb lattice with Kekule distortion, the $C_{6}$ rotational symmetry can be decomposed into combination of $C_{2}$ and $C_{3}$ rotational symmetry; the states at $\Gamma$, $M$ and $K$ points have $C_{6}$, $C_{2}$ and $C_{3}$ symmetry, respectively. Thus, we define the symmetric invariant of the Floquet band gap around the $\omega/2$ quasi-energy level at $M$ and $K$ points as
\begin{equation}
[M_{1}^{(2)}]\equiv \sharp M_{1}^{(2)}-\sharp \Gamma_{1}^{(2)}
\end{equation}
and
\begin{equation}
[K_{1}^{(3)}]\equiv\sharp K_{1}^{(3)}-\sharp \Gamma_{1}^{(3)}
\end{equation}
where $\sharp \Pi_{p}^{(n)}$ is the number of states at $\Omega$ point below $\omega/2$ quasi-energy level with eigenvalue being $\Pi_{p}^{(n)}$. The Floquet topological crystalline insulators (FTCIs) is characterized by the index $\chi^{(6)}=([M_{1}^{(2)}],[K_{1}^{(3)}])$ with nonzero corner charge. For the effective Hamiltonian with different $n_{Max}$, the number of bulk quasi-energy band below $\omega/2$ quasi-energy is different, i.e. is $3(2n_{Max}+1)+3$. However, as long as $n_{Max}\ge2$ is satisfied, the topological invariant is independent on $n_{Max}$.

The bulk band gap at high symmetric points and the corresponding topological invariant of two types of systems with $d_{1}=0.1d_{0}$ and $d_{1}=-0.1d_{0}$ are calculated and plotted in top and bottom rows in Fig. \ref{figure_kekule_bulkTopo}. The amplitude of the dynamic Kekule factor is $d_{2}=0.15d_{0}$ for both cases. The quasi-energy spectrum for the corresponding finite waveguide arrays are plotted in Fig. \ref{figure_kekule_band}(a,b) and (c,d), respectively. In some regime of $\omega$, the bulk band gap at the high symmetric points are sizable, but the quasi-energy spectrum is gapless around the $\omega/2$ quasi-energy level, because indirect band gap of the bulk energy band is negative. For a particular state, the degree of localization at certain region $\Omega$ can be calculated as $\sum_{i\in\Omega}{\rho_{i}}$. If $\Omega$ covers all lattice sites, the summation equates one. If $\Omega$ covers the lattice sites near to the two zigzag corners, the two armchair corners, or the four armchair edges, the summation equate to the degree of localization at the corresponding region. The size of the colored marker in Fig. \ref{figure_kekule_band}(a-d) is proportional to the corresponding degree of localization.

For the systems with $d_{1}=0.1d_{0}$, within the region $\omega/c_{0}\in[2.78,5.94]$, the bulk energy band of the 0-th and 1-st replicas overlap at quasi-energy level $\omega/2$. The coupling between the two replicas induce a Floquet energy gap of bulk states around the $\omega/2$ quasi-energy level, as shown in Fig. \ref{figure_kekule_bulkTopo}(a). The topological invariant in the dynamic bulk gap is $\chi^{(6)}=(2,0)$, as shown in Fig. \ref{figure_kekule_bulkTopo}(b). The corner charge of the bulk gap is $mod(\frac{1}{4}[M_{1}^{(2)}]+\frac{1}{6}[K_{1}^{(32)}],1)=\frac{1}{2}$, which imply that two corner states with $\omega/2$ quasi-energy level should appear \cite{Wladimir19}. The corner states are designated as $\pi$ corner states. However, multiple edge bands appear within this gap, as shown by the line with green markers in Fig. \ref{figure_kekule_band}(a,b). The edge band interfere with the $\pi$ corner states, so that some of the $\pi$ corner states are delocalized. Within the region $\omega/c_{0}\in[4.05,5.70]$, the energy spectrum of the edge states do not overlap with $\omega/2$, so that the $\pi$ corner states remain being localized. The spatial distribution of the magnitude of a typical $\pi$ corner states is plotted in Fig. \ref{figure_kekule_band}(e), which shows that the maximum of the mode magnitude is at the two sites that are nearest neighboring to the zigzag termination. Within the region $\omega/c_{0}\in[2.45,2.78]$, the topological invariant is $\chi^{(6)}=(4,0)$, and then the corner charge is zero, so that there is no $\pi$ corner state. Within the region $\omega/c_{0}<2.45$, the topological invariant is $\chi^{(6)}=(5,0)$, and then the corner charge is $\frac{1}{4}$, so that four corner states should appear. However, the indirect bulk band gap is negative, so that the interference between the corner states and the bulk states delocalized the corner states. In addition to the $\pi$ corner states, multiple corner states that localized at the zigzag corners or the armchair corners with $\varepsilon/\omega$ being dependent on $\omega/c_{0}$ appear. These corner states are due to the interfere of multiple bulk states in finite size systems, so that they are not topological. Some of the corner states are localized at the armchair corners. The spatial distribution of the magnitude of a typical armchair corner state is plotted in Fig. \ref{figure_kekule_band}(f).

The zero energy corner states are also modified by the dynamic modulation. With $\omega/c_{0}\in[1.60,2.15]$ or $\omega/c_{0}\in[2.55,2.95]$, the bulk energy band of the $\pm1$-st replicas overlap at zero energy. The coupling between the $\pm1$-st replicas and the zero energy corner state delocalized the zero energy corner state, so that a Floquet band gap without zero energy corner state is opened. Within the region of $\omega/c_{0}\in[2.15,2.55]$ or $\omega/c_{0}\in[2.95,3.7]$, the overlap between the $\pm1$-st replicas is uncomplete, and then the coupling is weak, so that the bulk states remain gapless at zero energy. When $\omega/c_{0}>3.7$, the zero energy corner state does not overlap with the $\pm1$-th replicas of bulk state, so that it remains being localized.

When the static Kekule factor flips sign, i.e., $d_{1}=-0.1d_{0}$, the bulk band gap, topological invariant of the Floquet band gap and the quasi-energy spectrum for the corresponding finite waveguide arrays are changed. The static model is topologically trivial \cite{Wladimir22}, so that the zero energy corner state does not appear with varying $\omega/c_{0}$. Within the region $\omega/c_{0}\in[2.32,2.84]$, the topological invariant is $\chi^{(6)}=(2,0)$, and then the corner charge is $\frac{1}{2}$, so that there are two $\pi$ corner states, which is confirmed by the numerical result in Fig. \ref{figure_kekule_band}(c,d). In addition, the bands associated with armchair corner states are located within the gap between the $\omega/2$ quasi-energy and the quasi-energy of the bulk bands, as shown by the lines with blue markers. Within the region $\omega/c_{0}\in[2.84,4.17]$, the topological invariant is $\chi^{(6)}=(1,0)$, and then the corner charge is $\frac{1}{4}$, so that there should be four corner states. However, the quasi-energy spectrum of the bulk states is near to $\omega/2$ quasi-energy level, which delocalized some of the corner states. Within a part of this region (specifically, $\omega/c_{0}\in[2.84,3.55]$), two corner states at the zigzag corners with $\omega/2$ quasi-energy level are found in the numerical results. Within the region $\omega/c_{0}\in[4.17,5.24]$, the topological invariant is $\chi^{(6)}=(0,0)$, and then the corner charge is zero, so that there is no corner state, as confirmed in Fig. \ref{figure_kekule_band}(c,d). Within the region $\omega/c_{0}<2.32$ or $\omega/c_{0}>5.24$, the indirect band gap of bulk is zero, so that the quasi-energy spectrum is gapless without corner state. Briefly, the zero energy corner states and $\pi$ corner states at the zigzag corners can be controlled by the Floquet engineering.

\subsection{Breathing square lattice}

\begin{figure*}[tbp]
\scalebox{0.73}{\includegraphics{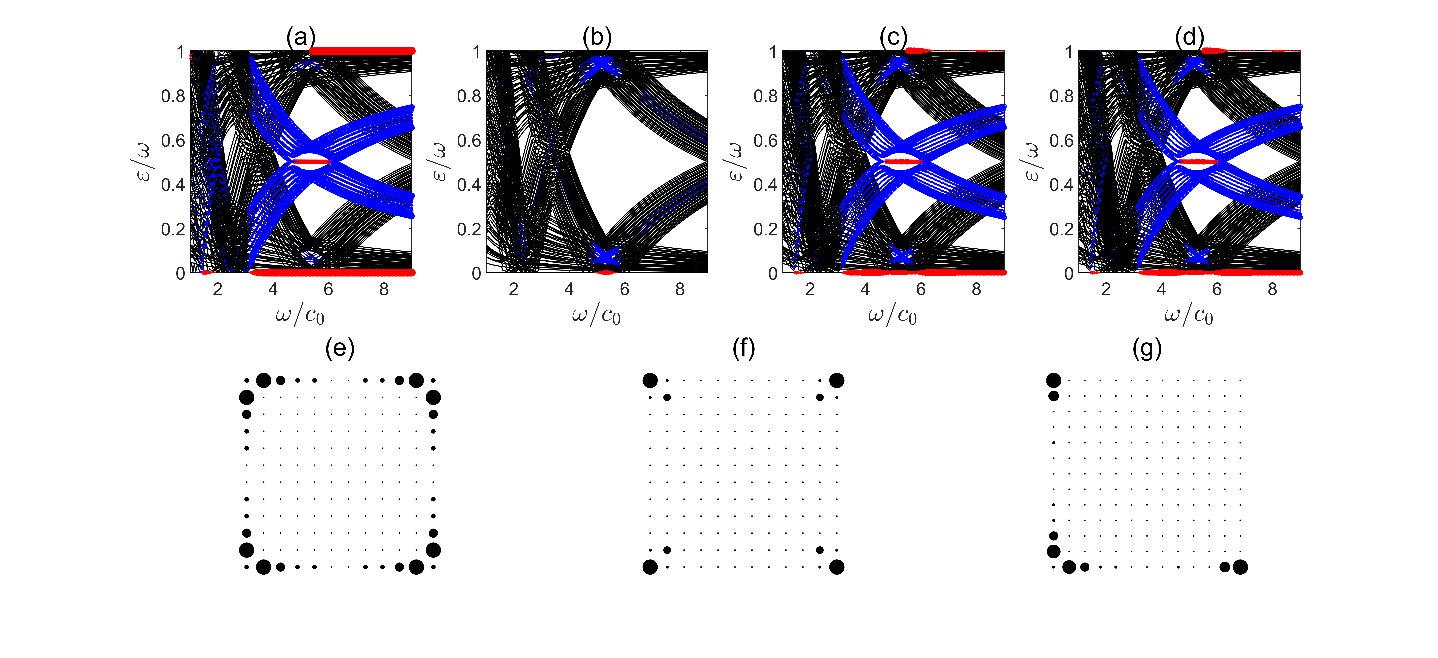}}
\caption{ Quasi-energy spectrum of the breathing square lattice in Fig. \ref{figure_system_kekule}(b) with $N=10$ for (a,b), $N=10.5$ for (c,d), as a function of $\omega/c_{0}$. The static breathing factors for the systems in (a,c) and (b,d) are $d_{1}=0.3d_{0}/\sqrt{2}$ and $d_{1}=-0.3d_{0}/\sqrt{2}$, respectively. The oscillating breathing factors for all systems are $d_{2}=0.2d_{0}/\sqrt{2}$. The states that are localized at the corners and edges are marked by red (circle) dots and blue (square) dots, respectively. The size of the markers is proportional to the degree of localization, i.e., $\sum_{i\in\Omega}\rho_{i}$. (e) and (g) are the spatial distribution of the mode amplitude of the $\pi$ modes of the systems in (a) and (c), respectively. (f) is the spatial distribution of the mode amplitude of the zero energy corner states of the systems in (a). }
\label{figure_square_band}
\end{figure*}

\begin{figure}[tbp]
\scalebox{0.58}{\includegraphics{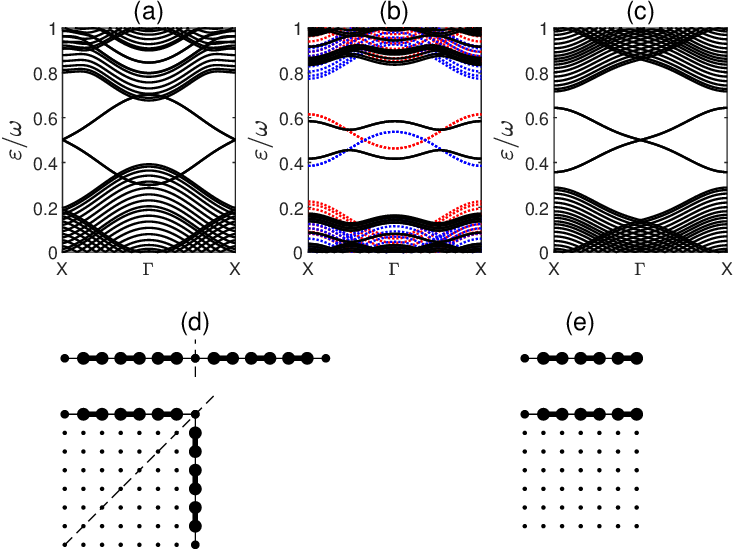}}
\caption{ (a-c) Quasi-energy band structure of the nanoribbon of the breathing square lattice in Fig. \ref{figure_system_kekule}(b) with $N=10$ in the width direction, $d_{1}=0.3d_{0}/\sqrt{2}$, and $d_{2}=0.2d_{0}/\sqrt{2}$. The axis direction of the nanoribbon is along the x axis. The oscillating frequency is $\omega/c_{0}=4.44$ in (a), $5.34$ in (b), and $6.235$ in (c). (d) Demonstration of the Floquet SSH edge chain around the 90$^{o}$ corner of the square lattice when $2N$ is even number. The axis of the reflection symmetric along the diagonal direction of the corner is plotted as dashed line. The equivalent 1D Floquet SSH chain with domain wall (plotted as dashed line) is plotted above the square lattice. (e) Demonstration of the Floquet SSH edge chain at the corner between topological and trivial edges of the square lattice with $2N$ being odd integer. The equivalent 1D Floquet SSH chain with open boundary is plotted above the square lattice. }
\label{figure_square_bandbulk}
\end{figure}

For the waveguide arrays in breathing square lattice, the tight binding Hamiltonian is given as
\begin{eqnarray}
&&H_{n}=\\
&&\begin{bmatrix}
0 & c_{n,1}+c_{n,2}^{(1)} & 0 & c_{n,1}+c_{n,2}^{(2)} \\
c_{n,1}+c_{n,2}^{(1)*} & 0 & c_{n,1}+c_{n,2}^{(2)} & 0 \\
0 & c_{n,1}+c_{n,2}^{(2)*} & 0 & c_{n,1}+c_{n,2}^{(1)*} \\
c_{n,1}+c_{n,2}^{(2)*} & 0 & c_{n,1}+c_{n,2}^{(1)*} & 0
\end{bmatrix}\nonumber
\end{eqnarray}
where $c_{n,2}^{(1)}=c_{n,2}e^{i\mathbf{k}\cdot\mathbf{a}_{1}}$, $c_{n,2}^{(2)}=c_{n,2}e^{i\mathbf{k}\cdot\mathbf{a}_{2}}$, $\mathbf{k}$ is the Bloch wave vector, $c_{n,1}=\int_{0}^{L}{c_{0}exp\{-\frac{[d_{1}+d_{2}\sin(2\pi z/L)]\sqrt{2}}{\delta}-\frac{i2\pi nz}{L}\}dz}$, and $c_{n,2}=\int_{0}^{L}{c_{0}exp\{-\frac{[-d_{1}-d_{2}\sin(2\pi z/L)]\sqrt{2}}{\delta}-\frac{i2\pi nz}{L}\}dz}$. In the numerical calculation, $d_{0}\sqrt{2}/\delta=3$ is assumed. The indices of lattice site in one unit cell is sorted by counter-clockwise order. Assuming $n_{Max}=2$, the effective Floquet Hamiltonian is similar to that in Eq. (\ref{effHamiHoney}). The modulation of the structures also preserve the $C_{4}$ rotational symmetric of the static breathing square lattice. The topological invariant of the crystalline symmetry can be defined, but the topological invariant of the Floquet band gap around the $\omega/2$ quasi-energy level is zero for this particular model. Thus, the $\pi$ corner states are generated by different mechanism.

The simulation results of the finite waveguide arrays are summarized in Fig. \ref{figure_square_band}. For the static system with $d_{2}=0$, each individual chain along $x$ or $y$ direction is an SSH model. As $d_{1}$ being positive, the SSH model is topological with two zero energy states that are localized at both ends. When the chains are coupled to the neighboring chains to form the breathing square lattice, four bands of bulk states are form. Assuming that the static intra-cell and inter-cell coupling strength are $c_{01}=c_{0}e^{(d_{0}-d_{1})\sqrt{2}/\delta}$ and $c_{02}=c_{0}e^{(d_{0}+d_{1})\sqrt{2}/\delta}$, the band structure of the four bulk bands are $E=\pm\sqrt{2}\sqrt{\pm\sqrt{c_{0x}c_{0y}}+c_{0xy}}$ with $c_{0x}=c_{01}^{2}+c_{02}^{2}+2c_{01}c_{02}\cos{k_{x}}$, $c_{0y}=c_{01}^{2}+c_{02}^{2}+2c_{01}c_{02}\cos{k_{y}}$ and $c_{0xy}=c_{01}^{2}+c_{02}^{2}+c_{01}c_{02}(\cos{k_{x}}+\cos{k_{y}})$,
$k_{x(y)}\in[-\pi,\pi]$ being the Bloch phase along the x(y) direction. At the high symmetric lines with $k_{x}=\pm k_{y}$, two bands (designated as band II and III) touch each other at zero energy, which form gapless energy spectrum around zero energy for the finite lattice array; two other bands (designated as band I and IV) are gapped from the former two bands with higher and lower energy level. Although the bulk band structure is gapless at zero energy, the zero energy corner states still exist in a finite lattice array. At zero energy, the dispersive of the two bulk bands are $k_{x}=\pm k_{y}$. In a square finite lattice array, the superposition of the zero energy bulk states forms four corner states and $N-4$ delocalized states. Thus, the four corner states are not topological. The end state at the end of each individual chain is coupled with the end states at the neighboring chains with alternating strength. The coupling end states form edge bands, which mimic a one dimensional edge SSH model along each edge. The edge SSH model has two bands (designated as SSH edge states), whose energy spectrum lay within the bulk gap between band III and IV (or band I and II). The propagating SSH edge states in each edge are bounded back and forth between two corners, which form standing waves with quasi-continue energy spectrum in a finite lattice array.

As the dynamic breathing factor $d_{2}$ being turned on, the SSH edge states of different Floquet replica couple to each other to form Floquet SSH edge states. When $\omega$ is near to the energy difference between the two bands of the SSH edge states, the coupling between the 0-th and $1$-st Floquet replica is large. In this case, a Floquet band gap around the quasi-energy level $\omega/2$ is opened. For varying $\omega$, the band structures of nanoribbon with $N=10$ unit cell along the width direction are plotted in Fig. \ref{figure_square_bandbulk}(a-c). At the critical value of $\omega/c_{0}=4.44$ and $\omega/c_{0}=6.235$, the band gap of the Floquet SSH edge states close at $X$ and $\Gamma$ point as shown in Fig. \ref{figure_square_bandbulk}(a) and (c), respectively. With $\omega$ being between the two critical values, the Floquet SSH edge states can be modeled as two one-dimensional (1D) massive Dirac Fermion modes. A typical example is exhibited in Fig. \ref{figure_square_bandbulk}(b) with maximum Floquet band gap. In the absence of the dynamic modulation (i.e., $d_{2}=0$), the bands of the 0-th and $1$-st Floquet replica of the SSH edge states cross at two points. Notice that the $1$-st Floquet replica is artificial for the system without modulation, so that the band crossing is also artificially imaginary. At the band crossing points, the band dispersion is nearly linear, so that the states can be modeled as 1D massless Dirac Fermion mode. By tuning on the modulation (i.e., $d_{2}\ne0$), the 0-th and $1$-st Floquet replica couple with each other to form the Floquet SSH edge states, so that the band crossing is not artificial. An effective mass term is generated for each Dirac Fermion model. The Floquet SSH edge states are localized near to the edge, so that they can be equivalent to a 1D Floquet SSH chain along the edge, as shown by lower part of Fig. \ref{figure_square_bandbulk}(d). At the corner of the square lattice, the Floquet SSH chain turns 90$^{o}$. Because the lattice structure have mirror symmetric about the diagonal direction, two consecutive bounds at the corner have the same strength. Thus, the Floquet SSH chain is equivalent to a 1D Floquet SSH chain with a domain wall, as shown by upper part of Fig. \ref{figure_square_bandbulk}(d). The masses of the two 1D massive Dirac Fermion modes flip sign at the domain wall, so that there are two corner states at the corner, which is protected by the mirror symmetric. The quasi-energy level of the corner states is $\omega/2$, as indicated by the line with red markers in Fig. \ref{figure_square_band}(a) at energy $\varepsilon/\omega=1/2$ and $\omega/c_{0}\in[4.8,6.05]$, so that the corner states are designated as $\pi$ corner states. The finite lattice array of the breathing square lattice has four corners, so that eight localized $\pi$ corner states appear at the quasi-energy level $\omega/2$. The spatial distribution of magnitude of the $\pi$ corner mode is plotted in Fig. \ref{figure_square_band}(e). The maximum magnitude is located at two lattice sites that is nearest neighboring to the terminate corner. As the $1$-st Floquet replica of the SSH edge states couple with the zero energy corner states, the zero energy corner state is delocalized. As shown in Fig. \ref{figure_square_band}(a), with $\omega/c_{0}\in[2,3.2]$, the zero energy corner states are delocalized.

In the opposite case when $d_{1}$ is negative, the bulk band structure remains being the same. However, the SSH model of each individual chain is topologically trivial. Thus, for the static system with $d_{2}=0$, the SSH edge state does not exist. The $N$ zero energy states are all delocalized bulk states. As the dynamic breathing factor $d_{2}$ being turned on, the coupling between the 0-th and 1-st Floquet replica of the the SSH edge states is absent due to the absence of the SSH edge states, so that the $\pi$ corner mode does not exist, as shown in Fig. \ref{figure_square_band}(b). The coupling between the 0-th Floquet replica of the band II/III and the $\pm1$-st Floquet replica of the band I/IV induces Floquet zero energy corner states, as the dynamic frequency is within the region $\omega/c_{0}\in[4.85,5.65]$, as shown in Fig. \ref{figure_square_band}(b). The Floquet zero energy corner states coexist with the gapless bulk states near zero energy. The spatial distribution of magnitude of the Floquet zero energy corner mode is plotted in Fig. \ref{figure_square_band}(f), which has the same pattern as the zero energy corner states in the static system with $d_{1}>0$ and $d_{2}=0$. The maximum magnitude is located at the terminate lattice site of each corner.

If the number of primitive unit cell along $x$ and $y$ direction is an integer plus $\frac{1}{2}$ (the number of lattice sites in the 1D chain is an odd number), the numerical result of the quasi-energy spectrum is shown in Fig. \ref{figure_square_band}(c) and (d). If a chain of the SSH model has odd number of sites, one of the end always host a topological end state, while the another end is trivial, because the nearest neighboring bonding to the terminations at the two ends have different strength. Similarly, if the breathing square lattice has odd number of lattice sites in each individual chain along $x$ and $y$ direction, two edges around a corner have SSH edge states, while the other two edges do not have SSH edge states. At the corner between the two topological edges with SSH edge states, two $\pi$ corner states are induced by the dynamical modulation. At a corner between a topological edge and a trivial edge, the Floquet SSH chain at the topological edge can be equivalent to a 1D Floquet SSH chain with trivial open boundary, as shown in Fig. \ref{figure_square_bandbulk}(e). Although the corner does not have mirror symmetry, one $\pi$ corner states appears due to the interfere between the two massive Dirac Fermion modes in the trivial open boundary of the 1D Floquet SSH chain. This corner state is not topological, because it is not protected by symmetry. The energy spectrum with the corner states are indicated by the red lines in Fig. \ref{figure_square_band}(c) and (d) with energy $\varepsilon/\omega=1/2$ and $\omega/c_{0}\in[4.8,6.05]$. The spatial distribution of the magnitude of the $\pi$ corner states are plotted in Fig. \ref{figure_square_band}(g) for the systems in Fig. \ref{figure_square_band}(c). In this case, the left and bottom edges are topological, so that two $\pi$ corners states are localized at the left-bottom corner, and one $\pi$ corner state are localized at each of left-top and right-bottom corners. For the $\pi$ corner states that locate at the corner between two topological edges, the maximum magnitude is located at two lattice sites that is nearest neighbor to the corner termination; for the $\pi$ corner states that locate at the corner between the topological edge and the trivial edge, maximum magnitude is located at the terminate lattice site of the corner, as shown in Fig. \ref{figure_square_band}(g).

\subsection{Breathing Kagome lattice}

For the waveguide arrays in breathing Kagome lattice, the tight binding Hamiltonian is given as
\begin{equation}
H_{n}=
\begin{bmatrix}
0 & c_{n,1}+c_{n,2}^{(2)} & c_{n,1}+c_{n,2}^{(3)} \\
c_{n,1}+c_{n,2}^{(2)*} & 0 & c_{n,1}+c_{n,2}^{(1)*} \\
c_{n,1}+c_{n,2}^{(3)*} & c_{n,1}+c_{n,2}^{(1)} & 0
\end{bmatrix}
\end{equation}
where $c_{n,2}^{(1)}=c_{n,2}e^{i\mathbf{k}\cdot\mathbf{a}_{1}}$, $c_{n,2}^{(2)}=c_{n,2}e^{i\mathbf{k}\cdot\mathbf{a}_{2}}$, $c_{n,2}^{(3)}=c_{n,2}e^{i\mathbf{k}\cdot(\mathbf{a}_{2}-\mathbf{a}_{1})}$, $\mathbf{k}$ is the Bloch wave vector, $c_{n,1}=\int_{0}^{L}{c_{0}exp\{-\frac{[d_{1}+d_{2}\sin(2\pi z/L)]\sqrt{3}}{\delta}-\frac{i2\pi nz}{L}\}dz}$, and $c_{n,2}=\int_{0}^{L}{c_{0}exp\{-\frac{[-d_{1}-d_{2}\sin(2\pi z/L)]\sqrt{3}}{\delta}-\frac{i2\pi nz}{L}\}dz}$. In the numerical calculation, $d_{0}\sqrt{3}/\delta=3$ is assumed. The indices of lattice site in one unit cell is sorted by counter-clockwise order. Assuming $n_{Max}=2$, the effective Floquet Hamiltonian is similar to that in Eq. (\ref{effHamiHoney}). The modulation of the structures also preserve the $C_{3}$ rotational symmetric of the static breathing Kagome lattice. Similar to that in subsection IIIA, the topological invariant of the crystalline symmetry at each Floquet band gap is defined as $[K_{2}^{(3)}]$, and the corner charge is $mod(\frac{1}{3}[K_{2}^{(3)}],1)$. For Kagome lattice, the flat band always stick to another dispersive band, so that the topological invariant between these two bands is not well defined.

\begin{figure*}[tbp]
\scalebox{0.73}{\includegraphics{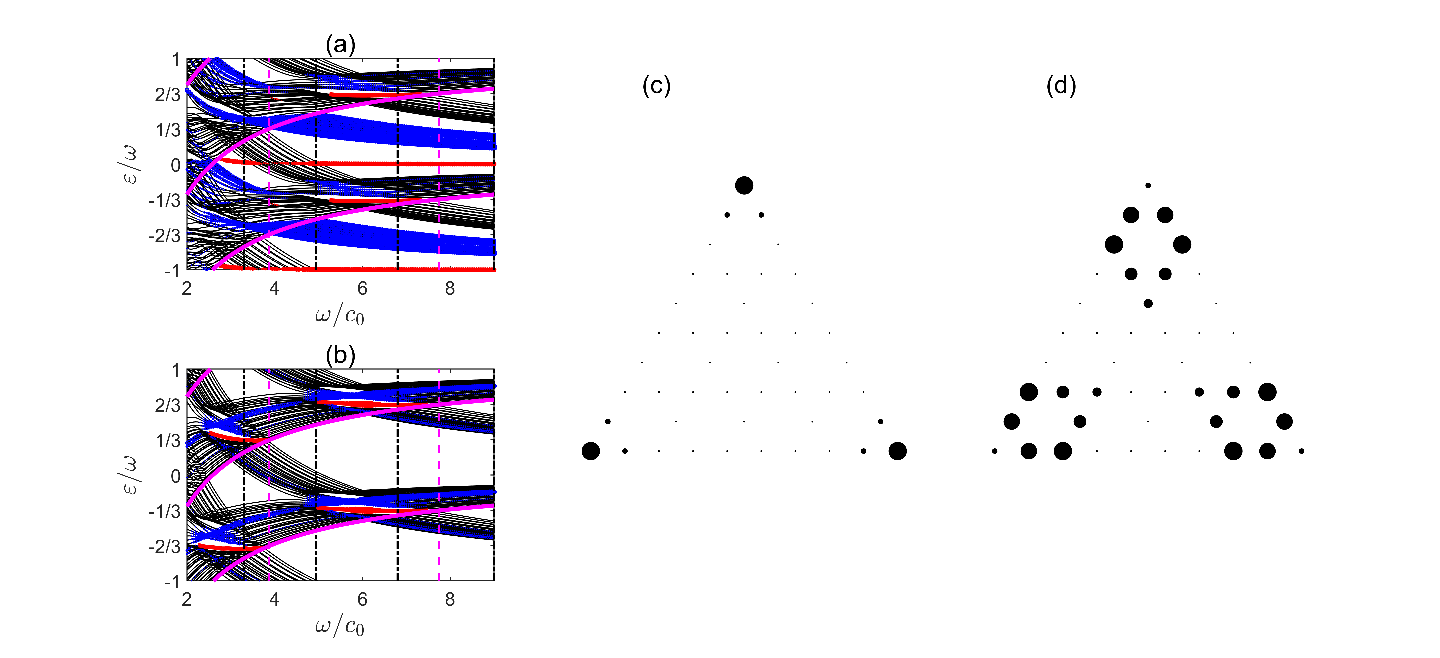}}
\caption{ (a,b) Quasi-energy spectrum of the breathing kagome lattice in Fig. \ref{figure_system_kekule}(c) with $N=10$, as a function of $\omega/c_{0}$, are plotted as black (solid) lines. The static breathing factor is $d_{1}=0.2d_{0}/\sqrt{3}$ for (a), and $d_{1}=-0.2d_{0}/\sqrt{3}$ for (b). The oscillating breathing factors for both systems are $d_{2}=0.2d_{0}/\sqrt{3}$. The states that are localized at the corners and edges are marked by red (circle) dots and blue (square) dots, respectively. The size of the markers is proportional to the degree of localization, i.e., $\sum_{i\in\Omega}\rho_{i}$. The magenta (thick solid) lines mark the states of the flat bands. The vertical magenta (dashed) and black (dash-dotted) lines mark the $\omega$ that the corresponding bulk bands are plotted in Fig. \ref{figure_kagome_bandbulk}. The vertical magenta (dashed) lines mark the critical $\omega$ where the triple band crossing occur. (c) is the spatial distribution of the mode amplitude of the zero energy corner states and A-type $\pm2\pi/3$ corner states; (d) is the spatial distribution of the mode amplitude of the T-type $-2\pi/3$ corner states. For better visualization, $N=5$ is used in (c,d). }
\label{figure_kagome_band}
\end{figure*}

For the waveguide arrays in breathing Kagome lattice, the simulation results are summarized in Fig. \ref{figure_kagome_band}. The static Kagome lattice with $d_{1}>0$ and $d_{2}=0$ is in the second order topological insulator phase, which has zero energy corner states \cite{Ezawa181}. On the other hand, the static Kagome lattice with $d_{1}<0$ and $d_{2}=0$ is trivial insulator without zero energy corner state.

As the dynamic breathing factor $d_{2}$ becoming nonzero, the Floquet band gaps are opened in the quasi-energy spectrum, which could host corner states with different properties. For the systems with $d_{1}>0$, the zero energy corner states remain being robust, as shown in Fig. \ref{figure_kagome_band}(a) by the line with red markers near to zero energy. Within the region $\omega/c_{0}\in[3.4,5.05]$, the zero energy corner states of 0-th replica and the bulk states of 1-st replica with energy near to zero coexist, which imply that the two types of states do not couple to each other. Because of the absence of particle-hole symmetric, bulk energy band of either $+1$-th or $-1$-th replica overlap with zero energy, so that the Floquet coupling between $\pm1$-th replica at zero energy is absent. As a result, the zero energy corner state remain being localized. The spatial distribution of the mode amplitude of the zero energy corner state is plotted in Fig. \ref{figure_kagome_band}(c). The modes are strongly localized at the terminate lattice sites of the corner. In addition to the zero energy corner states, other corner states appear within the Floquet band gap, whose energy level are marked by the lines with red markers near to the energy level $\varepsilon/\omega=2/3$ and $-1/3$. The spatial distribution of mode amplitude of these corner states is plotted in Fig. \ref{figure_kagome_band}(d). The mode amplitude is not localized at the terminate lattice site of the corners, but at the four lattice sites in adjacent to the corner termination. Since these corner states are induced from a second order topological insulator phase, they are designated as T-type $4\pi/3$ and $-2\pi/3$ corner states, which have energy level near to $\varepsilon/\omega=2/3$ and $-1/3$, respectively. In addition to the corner states, the edge states, which are localized at the edge of the lattice, appear within the bulk energy gap.

One the other hand, for the systems with $d_{1}<0$, the zero energy corner state is absent for any value of $\omega$. Within the region $\omega/c_{0}\in[4.95,7.8]$, the corner states with energy level near to $\varepsilon/\omega=2/3$ and $-1/3$ appear. Since these corner state are anomalously induced from trivial insulator phase, they are designated as A-type $4\pi/3$ and $-2\pi/3$ corner states, which have energy level near to $\varepsilon/\omega=2/3$ and $-1/3$, respectively. Within another region $\omega/c_{0}\in[2.6,3.89]$, the corner states with energy level near to $\varepsilon/\omega=1/3$ and $-2/3$ appear, which are designated as A-type $2\pi/3$ and $-4\pi/3$ corner states. For the A-type corner states, the spatial distribution of the mode amplitude is the same as that in Fig. \ref{figure_kagome_band}(c), which means that the modes are strongly localized at the terminate lattice sites of the corner. For both T-type and A-type corner states, the energy levels are slightly dependent on $\omega$, which is due to the absence of particle-hole symmetric in the model Hamiltonian.  Because the first Floquet Brillouin zone is $(-\omega/2,\omega/2]$, the state with quasi-energy $\varepsilon$ is equivalent to the other state with quasi-energy $\varepsilon+n\omega$, so that the $\pm2\pi/3$ corner states is equivalent to the $\mp4\pi/3$ corner states. Thus, the systems in Fig. \ref{figure_kagome_band}(a) have one corner state, which is designated as T-type $-2\pi/3$ corner state; the systems in Fig. \ref{figure_kagome_band}(b) have two corner states, which are designated as A-type $\pm2\pi/3$ corner states.

\begin{figure}[tbp]
\scalebox{0.58}{\includegraphics{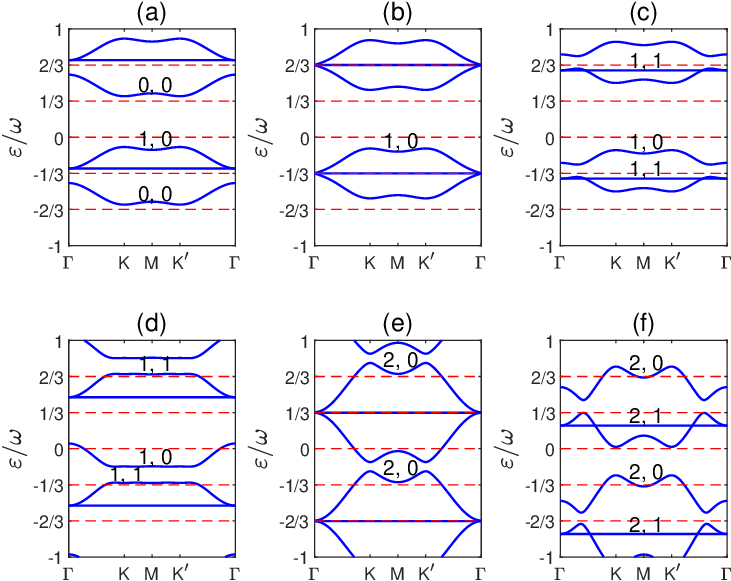}}
\caption{ Quasi-energy spectrum of the bulk state in breathing kagome lattice with the same parameters as those in Fig. \ref{figure_kagome_band}(a). $\omega/c_{0}$ is equal to (a) 9, (b) 7.8, (c) 6.8, (d) 4.95, (e) 3.89, (f) 3.3, respectively. The topological invariant of the Floquet band gap, i.e., $[K_{2}^{(3)}]$, are marked within the gap. For each band gap, the first and second values of $[K_{2}^{(3)}]$ are for the system with $d_{1}=0.2d_{0}/\sqrt{3}$ and $d_{1}=-0.2d_{0}/\sqrt{3}$, respectively. In some case, two bands are stuck together without gap, so that the topological invariant is not defined.  }
\label{figure_kagome_bandbulk}
\end{figure}

As $\omega$ varying, the transition between the Floquet states with Floquet band gap that does or does not host the T(A)-type $\pm2\pi/3$ corner states is accompanied by the triple band crossing among three bulk bands, two of which are dispersive and one of which is flat. Before and after the triple band crossing, the middle band is stuck to the higher and lower bands, and is gapped from the lower and higher bands, respectively. The value of $[K_{2}^{(3)}]$ of the gap changes at the triple band crossing. The critical value of $\omega$ at two triple band crossings are marked by the vertical magenta (dashed) lines in Fig. \ref{figure_kagome_band}(a) and (b). For the two systems in Fig. \ref{figure_kagome_band}(a) and (b) with positive and negative $d_{1}$, the bulk band structures are the same, which are plotted in Fig. \ref{figure_kagome_bandbulk} for selected $\omega$. As $\omega=9 c_{0}$, the Floquet band gap does not host corner state near to energy level $\pm2/3\omega$ or $\pm1/3\omega$. In this case, the two flat bands' energy level are above $2/3\omega$ and $-1/3\omega$; they are attached with a dispersive bulk band with higher energy (designated as higher band) at $\Gamma$ point, and are dynamically gapped from the dispersive bulk bands with lower energy (designated as lower band), as shown in Fig. \ref{figure_kagome_bandbulk}(a). The topological invariant $[K_{2}^{(3)}]$ of the Floquet band gap around $2/3\omega$ and $-1/3\omega$ are zero for both systems. As $\omega$ being decreased to be $7.8c_{0}$, triple band crossing occur at the $\Gamma$ point. The triple band crossing occurs at quasi-energy level $\frac{2}{3}\omega$, which is equal to the energy level of the flat band (designated as $\varepsilon_{flat}$); the higher and lower bands attach to the flat band at the $\Gamma$ point, as shown in Fig. \ref{figure_kagome_bandbulk}(b). Because the state with quasi-energy $\varepsilon$ is equivalent to the other state with quasi-energy $\varepsilon+n\omega$, another triple band crossing at quasi-energy level $-\frac{1}{3}\omega$ is the same as the triple band crossing at quasi-energy level $\frac{2}{3}\omega$. The triple band crossings occur as long as $\varepsilon_{flat}$ is equal to $2/3\omega$. As $\omega$ being further decreased, $\varepsilon_{flat}$ become smaller than $2/3\omega$; the flat bands are gapped from the higher band; the lower band partially penetrates through the flat band so that part of the dispersive band near to the $\Gamma$ point is above $\varepsilon_{flat}$; $[K_{2}^{(3)}]$ of the Floquet band gap between the higher and lower bands equates one for both systems, as shown in Fig. \ref{figure_kagome_bandbulk}(c). Thus, the T-type and A-type $\pm2\pi/3$ corner states appear within the Floquet band gap between the higher and lower bands. As $\omega$ being decreased to $4.95c_{0}$, the lower band has completely penetrated through the flat band; part of the dispersive bands become flat, where the energy level overlap with the T-type and A-type corner states, as shown in Fig. \ref{figure_kagome_bandbulk}(d). Within the Floquet band gap, $[K_{2}^{(3)}]$ remain being one, so that the T(A)-type corner states should appear. However, the mixing between the T(A)-type corner states and the bulk state delocalized the T(A)-type corner states, so that the markers of the T(A)-type corner states in the quasi-energy spectrum in Fig. \ref{figure_kagome_band}(a) and (b) cease to exist in this region of $\omega$ (i.e. $\omega/c_{0}\in[3.89,4.95]$). As $\omega$ further decrease, the designation of the higher and lower bands of the flat band are shifted: the higher (lower) band of the flat band at $\varepsilon_{flat}-\omega$ ($\varepsilon_{flat}$) becomes the lower (higher) band of the flat band at $\varepsilon_{flat}$. As $\omega$ reaches $3.89c_{0}$, another triple band crossings occur, as $\varepsilon_{flat}$ being equal to $1/3\omega$, as shown in Fig. \ref{figure_kagome_bandbulk}(e). As $\omega$ being further decreased, the flat band is gapped from the higher bands again; $[K_{2}^{(3)}]$ of the Floquet band gap equates two and one for the systems with positive and negative $d_{1}$, respectively. For the systems with $d_{1}>0$, $[K_{2}^{(3)}]=2$, and then the corner charge is $\frac{2}{3}$, so that the corner states should exist. However, the energy spectrum of the edge states overlap with that of the corner state, so that the corner states are delocalized. By contrast, for the system with $d_{1}<0$, $[K_{2}^{(3)}]=1$, and then the corner charge is $\frac{1}{3}$, so that the gaps host A-type $2\pi/3$ corner state with energy level near to $1/3\omega$, as shown in Fig. \ref{figure_kagome_band}(b).

\section{Discussion and Conclusion}

The numerical results show that the dynamic model of the three types of photonic waveguide arrays have Floquet corner states with energy levels being $\pm\frac{1}{2}\omega$ or $\pm\frac{1}{3}\omega$. The honeycomb lattice with Kekula distortion and the breathing square lattice have $C_{6}$ and $C_{4}$ rotational symmetry, respectively, both of which contain $C_{2}$ rotational symmetry, so that the quasi-energy level of the corner state is $\omega$. On the other hand, the breathing Kagome lattice have $C_{3}$ rotational symmetry, which does not contain $C_{2}$ symmetry, so that the quasi-energy level of the corner state is integral multiple of $\omega/3$. The Floquet corner state with fraction-$\pi$ quasi-energy level could be applied to engineer varying type of photonic devices. The coupling between Floquet $\pm2\pi/3$ corner states and the other type of corner states could offer feasible candidate as observation of novel type of photonic state, such as time-crystalline phases and period-triple oscillations \cite{Kuzmiak20}, and as assistant medium for braiding of photonic states in quantum optical devices \cite{JihoNoh20}.

For the model in honeycomb lattice with periodic Kekule distortion and that in Kagome lattice with periodic breathing factor, the dynamic coupling generates Floquet band gap directly from the bulk states. As a result, for both positive and negative $d_{1}$, the dynamic coupling with nonzero $d_{2}$ at appropriate frequency could open the Floquet band gap that hosts the Floquet corner states. The Floquet band gap is due to the coupling between the bulk energy band of 0-th and 1-st replica. At certain region of $\omega$, the system is in the FTCIs phase. On the other hand, for the model in square lattice with periodic breathing factor, the Floquet band gap is generated from the coupling between 0-th and 1-st replicas of the band of the SSH edge states. For the case that each 1D chain are trivial SSH lattice, the SSH edge states is absent, so that the Floquet band gap and Floquet $\pi$ corner state is also absent.

Concerning the efficiency of generating localized corner states, the performance of the breathing Kagome lattice is better than that of the other two types of lattice. Only two types of Floquet corner states are highly localized at the terminate lattice site at the corner: the A-type corner states in the breathing Kagome lattice (as shown in Fig. \ref{figure_kagome_band}(d)), and the $\pi$ corner states in breathing square lattice with $N$ being integer plus $\frac{1}{2}$ (as shown in Fig. \ref{figure_square_band}(g)). The other types of Floquet corner states are localized at the lattice sites neighboring to the terminate lattice site. Meanwhile, among the three type of lattice structure, the Kagome lattice has the fewest lattice site in one unit cell. Thus, the breathing Kagome lattice with $d_{1}<0$ is the most suitable for the application that require higher localization. On the other hand, the energy levels of the corner states in the breathing Kagome lattice are not fixed to a certain value, but slightly varying around $\pm\frac{1}{3}\omega$. Thus, the other two lattice models are more suitable for the application that require constant energy level for varying $\omega$.

In conclusion, the Floquet corner states at the outer corner of the photonic waveguide arrays can be engineered by the periodic modulation of the coupling strength between neighboring waveguides. The coupling between the bulk band of $\pm1$-st replicas can delocalized the zero energy corner state. The coupling between the bulk bands of 0-th and 1-st replicas could induce Floquet band gap with Floquet corner states. For waveguide arrays in honeycomb lattice and square lattice, the Floquet corner states have $\pi$ quasi-energy level due to the presence of the particle-hole symmetric and $C_{2}$ rotational symmetric. For waveguide arrays in Kagome lattice, the particle-hole symmetric is absent, and the systems have $C_{3}$ rotational symmetric, so that the Floquet corner state have fractional-$\pi$ quasi-energy level; the transition between the Floquet gaps that does or does not host the T(A)-type corner states is accompanied by the triple band crossing. The Floquet corner states of the honeycomb lattice and the Kagome lattice are topological, which are protected by the crystalline symmetry. The Floquet corner states of the square lattice are either topological that is protected by reflection symmetry, or non-topological that is due to interfere between dispersive edge states. The Floquet corner states at the open corners of the waveguide arrays could be applied in integrated photonic systems that require strong localization as well as efficient coupling with outside environment.

\begin{acknowledgments}
This project is supported by the Natural Science Foundation of Guangdong Province of China (Grant No.
2022A1515011578), the Project of Educational Commission of Guangdong Province of China (Grant No. 2021KTSCX064), the startup grant at Guangdong Polytechnic Normal University (Grant No. 2021SDKYA117), and the National Natural Science Foundation of China (Grant No.
11704419).
\end{acknowledgments}

\clearpage

\end{document}